\documentclass[sigconf]{acmart}
\usepackage{tabularx}
\usepackage{array}
\usepackage{enumitem}
\usepackage{colortbl}
\usepackage{makecell} % in preamble
\usepackage{tabularx} % in preamble
\usepackage{soul}

 \renewcommand{\shortauthors}{Das et al.}

\rowcolors{2}{gray!15}{white}
\AtBeginDocument{%
  }

\copyrightyear{2026}
\acmYear{2026}
\setcopyright{cc}
\setcctype{by}
\acmConference[CHI '26]{Proceedings of the 2026 CHI Conference on Human Factors in Computing Systems}{April 13--17, 2026}{Barcelona, Spain}
\acmBooktitle{Proceedings of the 2026 CHI Conference on Human Factors in Computing Systems (CHI '26), April 13--17, 2026, Barcelona, Spain}
\acmPrice{}
\acmDOI{10.1145/3772318.3791916}
\acmISBN{979-8-4007-2278-3/2026/04}

\begin{document}

%%
%% The "title" command has an optional parameter,
%% allowing the author to define a "short title" to be used in page headers.

\title[Image Obfuscation Techniques to Enhance Sharing Privacy for BVI]{Are You Comfortable Sharing It?: Leveraging Image Obfuscation Techniques to Enhance Sharing Privacy for Blind and Visually Impaired Users}

%\title[I Feel Vulnerable Sharing What I Can't See]{``I Feel Vulnerable Sharing What I Can't See”: Exploring Image Obfuscation Techniques for Blind Visually Impaired \kh{question on title}}
% Safe Sharing: Mitigating Privacy Concerns for Blind Individuals in Image Sharing
% Sharing with Confidence: Image Obfuscation Methods for the Visually Impaired
%%
%% The "author" command and its associated commands are used to define
%% the authors and their affiliations.
%% Of note is the shared affiliation of the first two authors, and the
%% "authornote" and "authornotemark" commands
%% used to denote shared contribution to the research.
\author{Satabdi Das}
\affiliation{%
  \institution{The University of British Columbia}
  \city{British Columbia}
  \country{Canada}}
\email{das24@student.ubc.ca}

\author{Nahian Beente Firuj}
\affiliation{%
  \institution{Shahjalal University of Science and Technology}
  \city{Sylhet}
  \country{Bangladesh}}
\email{nahian93@student.sust.edu}

\author{Manjot Singh}
\affiliation{%
  \institution{The University of British Columbia}
  \city{British Columbia}
  \country{Canada}}
\email{mnjt.singh2000@gmail.com}

\author{Arshad Nasser}
\affiliation{%
  \institution{The University of British Columbia}
  \city{British Columbia}
  \country{Canada}}
\email{arshad.nasser@ubc.ca}

\author{Khalad Hasan}
\affiliation{%
  \institution{The University of British Columbia}
  \city{British Columbia}
  \country{Canada}}
\email{khalad.hasan@ubc.ca}

%%
%% By default, the full list of authors will be used in the page
%% headers. Often, this list is too long, and will overlap
%% other information printed in the page headers. This command allows
%% the author to define a more concise list
%% of authors' names for this purpose.
\renewcommand{\shortauthors}{Das et al.}

%%
%% The abstract is a short summary of the work to be presented in the
%% article.
\begin{abstract}
  % People with Blind and Visual Impairments (BVI) often encounter challenges when sharing images to access visual information, as these images may accidentally contain sensitive or unintended content. In many instances, they are unaware of the potential risks associated with sharing such content, which can compromise their privacy and interpersonal relationships. To mitigate these concerns, we explored a set of image obfuscation/filtering techniques that BVI individuals are interested in applying to sensitive images before sharing them with different groups, such as family members, friends, or strangers. We conducted a study with 20 BVI participants to evaluate the effectiveness of these techniques on different images varying in sensitivity (e.g., personal moments, embarrassing shots) in reducing potentially sensitive or inappropriate content in shared images. Results revealed that pixelation was the least preferred technique, while no other single filtering method emerged as universally preferred; instead, choices varied depending on the content type. Additionally, participants expressed greater comfort sharing filtered images with different audiences than non-filtered images. Based on the results, we offer a set of design guidelines to enhance the image-sharing experience for BVI individuals.
% \sat{
People with Blind Visual Impairments (BVI) face unique challenges when sharing images, as these may accidentally contain sensitive or inappropriate content. In many instances, they are unaware of the potential risks associated with sharing such content, which can compromise their privacy and interpersonal relationships. To address this issue, we investigated image filtering techniques that could help BVI users manage sensitive content before sharing with various audiences, including family, friends, or strangers. We conducted a study with 20 BVI participants, evaluating different filters applied to images varying in sensitivity, such as personal moments or embarrassing shots. Results indicated that pixelation was the least preferred method, while preferences for other filters varied depending on image type and sharing context. Additionally, participants reported greater comfort when sharing filtered versus unfiltered images across audiences. Based on the results, we offer a set of design guidelines to enhance the image-sharing experience for BVI individuals.
\end{abstract}

%%
%% The code below is generated by the tool at http://dl.acm.org/ccs.cfm.
%% Please copy and paste the code instead of the example below.
%%

%%
%% Keywords. The author(s) should pick words that accurately describe

%% the work being presented. Separate the keywords with commas.

%%
%% This command processes the author and affiliation and title
%% information and builds the first part of the formatted document.
\maketitle

% sections are each in separate files

%!TEX root = paper.tex
\section{Introduction}

% Paragraph 1: The problem context and motivation for the project idea (why it's useful, what problem it solves) 

% Paragraph 2: how your project idea is situated among the most relevant related work (similar systems, algorithms, or techniques) 

% Paragraph 3: What you did (e.g., in this paper, we explored the idea of ..... )

% Paragraph 4: study summary and your key findings

% Paragraph 5 with bullet points: list of 1 to 4 main contributions
% https://dl.acm.org/doi/pdf/10.1145/3373625.3417014
% https://dl.acm.org/doi/pdf/10.1145/3613904.3642713
% https://dl.acm.org/doi/pdf/10.1145/3555570
% https://dl.acm.org/doi/pdf/10.1145/2702123.2702334
% https://dl.acm.org/doi/pdf/10.1145/2470654.2481291

Blind and Visually Impaired (BVI) individuals often share photos with others to access visual information - such as by requesting descriptions of the image content from sighted peers or volunteers to understand objects, people, text, or scenes depicted in the photos \cite{zhao2017effect,bennett2018teens,10.1145/3663548.3675599}. They also use images to socialize, for example, by sharing moments from their lives with friends and family, to preserve memories, and express themselves creatively \cite{kuriakose2022tools,morris2016most,voykinska2016blind}. In addition, mobile cameras and computer vision enable `visually aware' applications to assist BVI with everyday tasks, such as identifying objects, recognizing faces, and reading text \cite{akter2020uncomfortable}. 
For instance, they utilize Visual Description Services (VDS) via phone cameras, which leverage technologies such as computer vision and artificial intelligence to perform tasks like recognizing objects, reading text, and identifying facial expressions. 
VDS fall into AI-powered (e.g., Seeing AI \cite{SeeingAI}), human-powered (e.g., Be My Eyes \cite{BeMyEyes}), and hybrid approaches that combine both (e.g., VizWiz \cite{10.1145/1866029.1866080}).
%Despite the benefits of VDS, it raises privacy concerns, including risks like identity theft through visible personal identifiers like faces, names, or documents in the background, unauthorized sharing of personal information, and exposure of private spaces. 
However, this convenience introduces privacy challenges, as BVI users may unintentionally capture and share images containing sensitive or inappropriate content (e.g., private documents, identifiable individuals, or revealing backgrounds) without being aware of it. As a result, their privacy may be compromised, such as by inadvertently disclosing personal or third-party information, and this can potentially strain interpersonal relationships due to perceived breaches of trust or unintended oversharing \cite{morris2016most,10.1145/3613904.3642030}. Addressing these privacy concerns is essential to empower BVI users to confidently and safely share photos without risking their own or others’ privacy.

Recent research has explored privacy-enhancing technologies to support blind individuals in managing the visual content they share \cite{alharbi2022understanding,stangl2023dump,10.1145/3663548.3675612}. A promising approach involves applying computer vision techniques to automatically detect and obscure sensitive image elements - such as through blurring, masking, or removal \cite{15,66,90,zhang2024designing,brunton2015obfuscation}. These studies highlight that blind users value having agency over how privacy is maintained in their photos and advocate for tools that are designed with accessibility in mind to support informed decision-making \cite{alharbi2022understanding,stangl2023dump,10.1145/3663548.3675635}. 
%Prior studies have begun exploring obfuscation tools for BVI users. \remove{For example, Alharbi et al. \cite{alharbi2022understanding} highlighted how automatic obfuscation often misaligns with blind users’ expectations, advocating for more user-driven approaches. Zhang et al. \cite{zhang2023imageally} introduced ImageAlly, a human-AI system that facilitates obfuscation with help from trusted allies. More recently, Zhang et al. \cite{zhang2024designing} developed a screen-reader-accessible prototype and conducted hands-on studies showing how design frictions and lack of contextual feedback affect trust and usability. However, these works primarily focused on narrow content categories or system design aspects. Our work complements and extends this research by systematically examining preferences across multiple obfuscation styles and evaluating comfort levels for sharing filtered content with different audience groups (e.g., friends, family, strangers).}
Although recent work has advanced privacy-preserving tools for BVI individuals, existing approaches reveal several gaps that motivate our work. Alharbi et al. \cite{alharbi2022understanding} explored speculative obfuscation in visual description services but did not compare different filtering techniques across sensitive content categories. Zhang et al. \cite{zhang2023imageally} proposed a human-AI hybrid workflow for detecting and redacting private information, yet focused on usability rather than comparing obfuscation across diverse content types. Similarly, prior work on computer-generated descriptions \cite{zhang2024designing} considered sense-making and trust but did not explore how obfuscation affects BVI users’ comfort sharing images with different audiences. As a result, the field lacks empirical evidence on how BVI individuals judge which image filters suit different sensitive content and how these choices affect their comfort in sharing.

Our research centers on two primary objectives:  i) understanding BVI individuals’ comfort preferences regarding image filtering techniques across various image categories, and ii) exploring their comfort levels with sharing unfiltered and filtered images with friends, family, and strangers. To address these objectives, we conducted a user study with 20 BVI participants, who expressed their comfort levels while sharing images with different audiences, such as friends, family, and strangers. Additionally, participants were presented with images containing potentially sensitive or embarrassing content and were given options for filtering, including blurring, pixelation, masking, and content-based filling techniques. Results from the study showed that no single technique fits all situations - preferences varied by image and content type. Additionally, participants felt more comfortable sharing filtered images with audiences than non-filtered images. These findings emphasize the need for customizable, context-aware filtering tools that assist BVI users in making confident sharing decisions. 
Rather than proposing new algorithms, our contributions articulate how system decisions should adapt to BVI users’ context-specific privacy expectations, filling a critical gap between technical capability and end-user comfort.

Our paper contributes to the field in the following ways:

\begin{itemize}

\item An investigation of the preferences of BVI individuals regarding image filtering techniques across different types of sensitive content.

 \item An examination how comfort levels in sharing images change for BVI users depending on the target audience before and after applying various image filters.

\item We present a set of design implications for image obfuscation techniques that are specifically tailored to the preferences of BVI individuals, supporting safer and more comfortable image sharing across different target audiences.

\end{itemize}

\section{Related Work}

% \outline{introduce the main topics you will cover, possibly explain why you aren't covering other topics; I usually leave the most relevant topic for last}

% \cite{buxton_touchgesturemarking_1995}

% \outline{summarise why your work is different than previous projects}

% \cite{Sridhar:2017:WatchSense}

\subsection{Automated Assistive Systems}
% https://dl.acm.org/doi/pdf/10.1145/3373625.3417014
% https://www.usenix.org/system/files/sec20-akter.pdf

Various camera-based assistive technologies, such as object identifiers, barcode readers, text readers, color readers, money readers, and crowd-sourced visual question-answering systems, have been developed to help BVIs with daily tasks like identifying objects, reading prescriptions, and answering questions \cite{kacorri2017people,4067752,akter2020uncomfortable,BeMyEyes,aira}.
The field of Human-Computer Interaction (HCI) has in recent years concentrated its research efforts on the intricate domain of private information within images \cite{f,k}, particularly those captured by individuals who are BVI \cite{66,j}. This line of investigation underscores the nuanced challenges confronted by the BVI community in harnessing visual assistive technologies effectively. One significant concern is the inadvertent capture of sensitive or confidential data within these images, leading to a notable hesitancy among BVI individuals when seeking visual assistance from sighted peers \cite{stangl2023dump}. This hesitancy emanates from a fear of privacy breaches and the unintended exposure of personal information. Consequently, there exists a potential curtailment of BVI individuals' engagement with the visual realm, a phenomenon that bears ramifications for their access to pertinent visual information and services. While existing assistive systems excel at recognizing objects, text, or scenes, there is limited support for BVI users in identifying whether an image contains sensitive content before sharing. Vision-based safety classifiers exist for sighted users (e.g., detecting nudity, document, or bystander), but these tools have not been adapted for non-visual workflows. As a result, BVI users lack tools to assess when an image violates social or contextual privacy norms, highlighting the need for tools to evaluate and manage sensitive content before sharing.
 
\subsection{Privacy Concerns of BVI}
% https://dl.acm.org/doi/pdf/10.1145/2702123.2702334
% https://dl.acm.org/doi/pdf/10.1145/2470654.2481291
% Challenges in Image Capture and Sharing for BVI Individuals
Recent advances in wearable and mobile computing improve accessibility for visually impaired people, yet their unique privacy and security needs are often overlooked. Tousif et al. \cite{Tousif} showed that while BVI individuals share many of the same high-level privacy concerns as sighted users - such as preventing unintended disclosure and avoiding eavesdropping - the severity and sources of these concerns differ. As BVI users cannot visually verify content and often rely on sighted assistance, they face unique risks of involuntary disclosure through others reading their documents or screen-reader audio leaks. Thus, their concerns overlap conceptually with sighted users but are intensified and shaped by accessibility barriers and systems not designed with their needs in mind. Yet no prior work examines how different privacy filters change BVI users' comfort or willingness to share, leaving unclear how to design tools that meaningfully reduce disclosure risk.
Within the HCI domain, scholarly endeavors have explored the intricacies surrounding the process of image capture by BVI individuals and their subsequent reliance on visual assistance mechanisms \cite{66,j}. The latent presence of private information within the spectrum of captured images emerges as a formidable impediment to effective assistance-seeking, especially when images are shared with sighted counterparts. Although select studies have undertaken the examination of privacy-related concerns \cite{66,stangl2023dump}, a discernible gap remains unaddressed. This gap pertains to methodological interventions aimed at systematically purging images of private content, thereby fostering a heightened sense of agency and empowerment within the BVI community in their utilization of visual assistance technologies. Despite identifying privacy risks, prior work does not examine how audience type (e.g., family, friends, strangers) affects BVI users' comfort when sharing sensitive images. Studies with sighted users showed that audience proximity strongly influences disclosure decisions; however, this has never been empirically tested with BVI users  \cite{koelle2020social}. This gap motivates our focus on audience-specific comfort levels and willingness to share.

\subsection{Accessible Obfuscation Design}
% https://dl.acm.org/doi/pdf/10.1145/3613904.3642713
% https://dl.acm.org/doi/pdf/10.1145/3555570
% Potential of Image Sanitization Techniques in Addressing Privacy Concerns
The advent of image sanitization techniques presents a promising avenue to mitigate the intricate privacy concerns that beset BVI individuals \cite{k,l}. These innovative techniques are designed to expunge private information from images, while concurrently safeguarding the essential visual content and its contextual coherence \cite{l,c}. Prior work has examined obfuscation methods—including blurring, pixelation, masking, and inpainting—as privacy-preserving tools for BVI users \cite{zhang2024designing, alharbi2022understanding}. Alharbi et al. \cite{alharbi2022understanding} conducted interviews revealing BVI users’ desire for greater control and the limitations of automatic filtering. Zhang et al. \cite{zhang2023imageally} introduced ImageAlly, a hybrid system allowing users to delegate obfuscation tasks to sighted allies, while maintaining agency. Zhang et al.\cite{zhang2024designing} prototyped a screen-reader-accessible interface and found that even with modern AI models, users face friction due to unclear system feedback and cognitive load. However, existing obfuscation research leaves several key gaps. First, most studies evaluated a single technique or narrow contexts (e.g., documents, faces), and rarely compared multiple obfuscation techniques across diverse sensitive content categories. Second, no prior work examined comfort for sharing filtered images across different audiences, despite BVI users sharing images with both intimate and unknown recipients. Third, prior evaluations focus primarily on system usability or workflow trust, leaving it unclear which obfuscation methods BVI users find comfortable for different types of content. Our study addresses these gaps by systematically varying content category, obfuscation method, and audience to understand how these factors shape comfort and sharing decisions.

% These methods encompass a spectrum of approaches, ranging from subtle blurring and meticulous masking to sophisticated content-based replacements \cite{c}. However, the extent to which these techniques succeed in preserving privacy while simultaneously rendering images accessible and meaningful for BVI individuals necessitates deeper exploration within the existing HCI scholarship.
Beyond the BVI community, privacy literature in adjacent domains offers useful but limited insights. For example, studies with sighted users in augmented reality and surveillance contexts showed that preferences for blurring, masking, or pixelation vary with content sensitivity and the anticipated social consequences of disclosure. Cruz et al.\cite{fi17020055} found that sighted users preferred masking for personal content and blurring for bystanders. However, these findings do not generalize to BVI users, who cannot visually verify filtered results and rely on descriptive feedback. Our study, therefore, provides the first BVI-centered examination of obfuscation preferences.

In the pursuit of enhancing the confidence of BVI individuals in seeking visual assistance, a pivotal imperative lies in developing specialized image filtering techniques that meticulously strike a balance between privacy safeguards and cognitive comprehension requisites. The chosen methodological framework should seamlessly excise private data, obscuring or substituting sensitive elements, all while preserving the underlying contextual fabric of the image. Moreover, these techniques should display adaptability across diverse image genres \cite{h,i}, thereby empowering BVI individuals to purify a broad spectrum of visual content effectively. 

We ground our work in Nissenbaum’s theory of contextual integrity \cite{nissenbaum2004privacy}, which views privacy as the appropriateness of information flows defined by actors, attributes, and transmission principles. Privacy violations occur when information reaches outside these contextual norms - a key concern for BVI users who cannot always see their photos. Our study investigates how obfuscation can adjust image attributes to align with audience expectations and maintain contextual integrity. We also draw on Privacy by Design principles \cite{cavoukian2009privacy}, emphasizing strong defaults, user agency, and accessible controls that support informed privacy decisions. By examining how content, filtering style, and audience interact, our work puts these theories into practice for BVI image-sharing scenarios.

 \section{Study}
We conducted a user study to explore the preferences of BVI individuals regarding the use of image obfuscation techniques across different image categories and target audiences. We examined participants’ comfort levels before and after applying filters, particularly in relation to the target audience. 
To avoid conflating related constructs, we explicitly defined how each was operationalized: \textit{comfort levels} captured participants' social-emotional ease or unease when imagining the image being shared with others and \textit{willingness to share} measured their stated likelihood of disclosure in typical social contexts. 
Participants rated their comfort level using a 7-point scale consistent with prior work on socially sensitive interactions \cite{rico2010gesture, ahlstrom2014you,10.1145/1851600.1851647,10.1145/2070481.2070551,10.1145/3411764.3445430,10.1145/3281505.3281541}, and were instructed to interpret ``comfort'' specifically in social-emotional terms, which may include factors like privacy, safety, and social norms.
Our research questions include:

\begin{itemize}
  \item \textbf{RQ1}. \textit{What are the preferred image filtering techniques (Blurring, Pixelation, Masking, and Content-Based filling) among BVI participants for different types of sensitive content (e.g., nudity, identity information)?}   
  
  Prior work with sighted users \cite{fi17020055} compared filtering methods e.g., blurring, pixelation, and masking, but focused on sighted preferences. Research involving BVI users \cite{alharbi2022understanding, zhang2024designing} emphasized system usability but covered only limited content types such as Identity information or medical details. Thus, little is known about how preference varies across a broader range of sensitive image categories (e.g., personal, medical, nudity). Addressing RQ1 allows us to examine how content sensitivity influences BVI users' filtering preferences.
  
  \item \textbf{RQ2}. \textit{How do comfort levels in sharing images vary across target audiences, i.e., family, friends, strangers, before and after applying image filters?} 
  Prior work on social acceptability \cite{koelle2020social} showed that the presence of audiences can influence users' interactions with technology. However, no prior research has examined how audience factors might affect photo-sharing comfort among BVI users. RQ2 therefore investigates the role of the audience as a potential moderating factor in privacy comfort.
  
\end{itemize}

\begin{table*}[t]
\centering
\resizebox{\textwidth}{!}{%
\begin{tabular}{p{0.05\linewidth}|p{0.2\linewidth}|p{0.75\linewidth}}
\hline
\textbf{Code} & \textbf{Category} & \textbf{Sensitive content examples} \\
\hline
C1 & The identity of people & 
\begin{tabular}[t]{@{}p{\linewidth}@{}}

- Two friends are standing, and a third person raises his leg toward their faces. The third person's leg is visible in the photo, not his face. \\
- One person is sitting in a scenic view, and another old person is visible on the side. \\
\end{tabular} \\
\hline
 C2& Children & 
\begin{tabular}[t]{@{}p{\linewidth}@{}}
- In a photo showcasing a scenic view of mountains and lakes, one child is also visible \\
- A person shows her dress, however, there are three kids in the background, one of whom is clearly visible in the picture. \\
\end{tabular} \\
\hline
C3 & Nudity or partial nudity & 
\begin{tabular}[t]{@{}p{\linewidth}@{}}
- On the beach, there is a naked man walking. \\
- An open magazine has three bike photos, each showing a clothed man and a naked woman. \\
\end{tabular} \\
\hline

C4 & Private area at home & 
\begin{tabular}[t]{@{}p{\linewidth}@{}}

- A picture showing a messy corner near a sofa with a woman’s undergarment. \\
- Four capsicums, near the sink desk, the sink is full of bad-looking food waste.
 \\
\end{tabular} \\
\hline
 C5& Embarrassing shots
 & 
\begin{tabular}[t]{@{}p{\linewidth}@{}}
- A friend is taking a selfie, and another friend sleeping with an open mouth. \\
- A picture shows two people sleeping inappropriately on train seats, with two pieces of luggage placed in front of them. \\
\end{tabular} \\
\hline
C6 & Identity information & 
\begin{tabular}[t]{@{}p{\linewidth}@{}}

-  A person is holding her credit card to pay at a store where all the credit card information is visible. \\
- A person is holding her passport and smiling, and the passport information page is open and visible.
 \\
\end{tabular} \\
\hline
 C7& Medical condition
 & 
\begin{tabular}[t]{@{}p{\linewidth}@{}}
- A doctor is doing an operation on a patient's leg, she is putting stitches in the patient's feet, which got a vertical cut, and blood is there. \\
-  A doctor is preparing an injection shot, and in front of her, there is a radiology report lying some of the text is readable.
 \\
\end{tabular} \\
\hline

C8
 & Habit/interest
 & 
\begin{tabular}[t]{@{}p{\linewidth}@{}}

-  A picture showing a person's collection of magazines - which shows his hobby, however, and there is a magazine cover with half-naked women, showing their buttocks.
 \\
- A person, who might explore satanic activities as a hobby—as suggested by the visible satanic symbols in the picture—is seen near a computer monitor.
 \\
\end{tabular} \\
\hline

C9
 & Illegal/inappropriate content
 & 
\begin{tabular}[t]{@{}p{\linewidth}@{}}

-   A book named Red Zone by Alan is on one side, and an Irish whiskey bottle is on the other side. \\
-  A mechanical keyboard and a drug test report are visible, the latter indicating that the individual tested positive for marijuana in their urine.
 \\
\end{tabular} \\
\hline
 
C10 & Personal moment
 & 
\begin{tabular}[t]{@{}p{\linewidth}@{}}

-   A person is watching a kissing photo of her friends on her phone.
\\
-  An aloe vera moisturizer, a couple cuddling intimately (the lady is on the lap of the guy) photo, and random paper on a surface.
\\
\end{tabular} \\
\hline

\end{tabular}
}
\caption{Image categories with example images}
\Description{A table titled "Sensitive Content Examples" categorizes types of sensitive content in images with examples for each. There are three columns: "Code", "Category", and "Sensitive content examples". The categories include:

C1: The identity of people (e.g., a leg is visible but not the face; an unexpected person appears in the scene),

C2: Children (e.g., children appear unexpectedly in scenic or outfit-focused photos),

C3: Nudity or partial nudity (e.g., a naked man on a beach; naked women in magazines),

C4: Private area at home (e.g., undergarments in messy corners; food waste in the sink),

C5: Embarrassing shots (e.g., people sleeping in awkward positions in the background),

C6: Identity information (e.g., visible credit card or passport details),

C7: Medical condition (e.g., a leg operation with visible blood; radiology reports in view),

C8: Habit/interest (e.g., magazines with half-naked women; visible satanic symbols),

C9: Illegal/inappropriate content (e.g., alcohol with a drug test report),

C10: Personal moment (e.g., a person viewing an intimate photo of friends, or couples cuddling).

Each row includes two specific examples of potentially sensitive content for that category.}
\label{tab:sensitive_content}
\end{table*}

\section{Method}
%\subsection{Design Factors}

BVI individuals often share images with others - such as family members, friends, or even strangers - to gain contextual information about their surroundings or specific objects. This is done through directly sending images to friends and family members or strangers through tools like Seeing AI \cite{SeeingAI}, Be My Eyes \cite{BeMyEyes}, and VizWiz \cite{10.1145/1866029.1866080}. The images they capture and share can belong to various categories, including personal items, private spaces, or scenes with other people. To preserve privacy when sharing such content, different image obfuscation and filtering techniques can be applied to reduce the visibility of potentially sensitive information. In the sections below, we explore three key design factors - \textit{Image Category}, \textit{Image Obfuscation/Filtering Techniques}, and \textit{Target Audience} - to better understand how to support privacy-preserving image sharing for BVI individuals.

\subsection{Image category}
In order to determine BVI individuals' preferences for obfuscation techniques, we developed an image dataset grounded in the framework proposed by Li et al. \cite{data}. 
To clarify our design choices, we derived image categories from established privacy and computer vision ethics taxonomies \cite{adams2007sharing, orekondy2017towards, hoyle2015sensitive, aura2006scanning, besmer2010moving}. Through iterative author review, redundancy removal, and validation with a BVI researcher, we identified ten recurring forms of sensitive visual content. A pilot session (N=3) confirmed that ten categories offered sufficient coverage while keeping the study feasible within one hour. To balance representativeness and study duration, we selected two examples per category. From an initial pool of approximately six images per category, we chose those that 1) clearly conveyed the sensitive attribute, 2) offered diversity, 3) minimized confounding cues, and 4) were approved by the BVI researcher as privacy-relevant. This ensured the final image set was empirically grounded, methodologically tractable, and aligned with real-world privacy scenarios \cite{66, orekondy2017towards}.
The categories are: \textit{The identity of people, Children, Nudity or partial nudity, Private area at home, Embarrassing shots, Identity information, Medical condition, Habit/interest, Illegal/inappropriate content, Personal moment}. Three sighted HCI researchers analyzed and described each image to ensure the visual content was accurately captured and described from a sighted perspective. Their descriptions were then cross-validated by one BVI HCI researcher involved in this project to ensure that they were accessible, meaningful, and aligned with the experiences and needs of BVI individuals. Here are the categories that can classify images based on sensitive content (Table \ref{tab:sensitive_content}).

\begin{table*}[t]
\resizebox{\textwidth}{!}{%
\begin{tabularx}{\textwidth}{
>{\centering\arraybackslash}m{0.6cm} | % PID (wider so center works)
>{\centering\arraybackslash}m{0.6cm} | % Age
>{\centering\arraybackslash}m{1.0cm} | % Gender
p{3.9cm} |                                   % Impairment
X |                              % Assistive Device
>{\centering\arraybackslash}m{1cm}  % Photos
}
\hline
\textbf{Pid} & \textbf{Age} & \textbf{Gender} & \textbf{\makecell{Impairment}} & \textbf{Assistive Device} & \textbf{\makecell{Took\\Photos}} \\
\hline
1  & 29 & M      & Limited ability to detect light & Screen reader                                                         & Yes  \\
2  & 31 & M      & Totally blind, from birth                                                                               & Screen reader                                                    & Yes  \\
3  & 23 & M      & Totally blind                       & NVDA, JAWS, VoiceOver, Orca                                     & Yes  \\
4  & 25 & M      & Totally blind                                       & InstaReader                                                            & Yes  \\
5  & 24 & F      & Limited ability to detect light                                       & InstaReader                                                            & Yes  \\
6  & 24 & F      &  Totally blind      & Screen reader, Eloquence, GShow, Samantha voice, TTS       & Yes  \\
7  & 25 & M      & Totally blind                                         & Eloquence, InstaReader, Auto TTS                                       & Yes  \\
8  & 25 & M      &  Totally blind, from birth                                        & Screen reader                                                          & No   \\
9  & 27 & F      & Totally blind, from birth      & Screen reader, Vision AI, Tech Freedom                                 & Yes  \\
10 & 25 & M      & Totally blind                                        & Android Accessibility Suite, GSU, Auto TTS                             & Yes  \\
11 & 23 & F      &  Limited ability to detect light & Auto TTS, InstaReader                                                  & Yes  \\
12 & 24 & F      & Limited ability to detect light  & Auto TTS, Screen reader, InstaReader                                   & Yes  \\
13 & 24 & M      & Limited ability to detect light & InstaReader, Screen reader, eSpeak, Envision AI, Kibo                & Yes  \\
14 & 25 & M      &Totally blind, from birth                                         & Auto TTS, InstaReader, Eloquence, Kibo                                 & Yes \\
15 & 20 & M     & Totally blind, from birth                                                  & Talkback, Auto TTS, Screen reader, Eloquence                                   & Yes\\
16 & 20 & M     & Totally blind                                    & Jieshuo, Auto TTS, Be My Eyes, Eloquence                                   & Yes\\
17 & 23 & F     & Totally blind                                 & Jieshuo                                  & Yes\\
18 & 21 & M     &  Limited ability to detect light                                        & Eloquence, Auto TTS, Screen Reader                                  & Yes\\
19 & 20 & M     &  Totally blind, from birth                                              & Eloquence, Auto TTS                                 & Yes\\
20 & 21 & F     &  Totally blind, from birth                                              & Eloquence, Be My Eyes, InstaReader                                & Yes\\
\hline
\end{tabularx}
}
\caption{Participant details}
\label{par}

\end{table*}
\subsection{Image obfuscation/filtering techniques}

Our study aimed to evaluate the effectiveness of image obfuscation techniques designed to conceal objects and their properties while maintaining the overall utility of the images. The obfuscation techniques included Blurring, Pixelation, Masking, and Content-based filling. We selected these four techniques as they are commonly used and studied in prior work \cite{zhang2024designing,ilia2015face,shu2018cardea,bragg2020exploring,padilla2015visual,hasan2018viewer}. To facilitate comprehension for BVI, we described each technique using metaphors. These metaphors were co-developed with a BVI author on the research team, who provided valuable insights to ensure the descriptions were accessible, meaningful, and grounded in non-visual experiences.

\begin{itemize}
    \item \textbf{Blurring}: Blur is the effect on specific areas of an image, making the details indistinguishable while retaining the general shapes and context. Blurring is useful for hiding sensitive information or faces while keeping the image's overall composition intact. For BVI individuals, we explained blurring as distorting the image: the shape is retained, but the colors get distorted, making it hard to tell what it is. Think of a voice on a call with a poor signal: you can tell someone is speaking and perhaps gauge their tone, but you can not understand the words.

    \item \textbf{Pixelation}: The technique converts the target area into a grid of large, indistinguishable pixels, concealing finer details. Pixelation is often used in media to censor explicit content. For BVI comprehension, we described it as similar to blurring, but with the shape becoming more distorted while most colors remain. It is like viewing a misty or cloudy image. Imagine a broken voice on a call: you might understand a few sounds or words here and there, but can not fully comprehend what is being said.
    
    \item \textbf{Masking}: Masking covers specific parts of an image with solid colors or patterns, effectively hiding the underlying content. This method is straightforward and can completely obscure sensitive areas without altering the surrounding image. For BVI individuals, we described masking as covering up private or sensitive elements in the image with a solid-colored shape. Think of the beep used to hide private information in phone call recordings.
    
    \item \textbf{Content-based Filling}: The technique uses algorithms to replace the obfuscated area with visually plausible content generated from the surrounding context. It aims to blend the obfuscated region with the rest of the image. For BVI individuals, we explained this as Artificial Intelligence (AI) technology that analyzes images and replaces private or sensitive content with elements from their surroundings. For example, if there is a person in a forest, the AI will remove the person and create a background based on the surrounding forest. It does not recreate reality but provides a good imitation of what might be behind the person. 
\end{itemize}
\begin{table*}[t]
\centering

\resizebox{\textwidth}{!}{%
\begin{tabular}{lccccc}
\toprule
\textbf{Code | Category} &\textbf{Friedman $\chi^2$ ($p$)} &  \textbf{Blurring} & \textbf{Pixelation} & \textbf{Content-Based Filling}&\textbf{Masking} \\
\midrule
C1 | The identity of people & 7.65 ($p=0.05$) & 2.50 & 2.50 & 4.00 & 4.00 \\
C2 | Children & 1.21 ($p=0.75$) & 1.00 & 2.25 & 4.00 & 1.25 \\
C3 | Nudity or partial nudity & 12.66 ($p=0.005$) & 5.25$^\ast$ & 2.50$^\ast$& 4.00 & 3.25 \\
C4 | Private area at home & 9.92 ($p=0.019$) & 4.00$^\ast$  & 2.50 $^\ast$ & 4.00 & 3.25 \\
 C5 | Embarrassing shots & 4.49 ($p=0.21$) & 5.00 & 3.25 & 4.00 & 4.00 \\
C6 | Identity information  & 14.60 ($p=0.002$) & 7.00$^\ast$ & 2.00$^\ast$ & 1.00 & 2.50 \\

 C7 | Medical condition & 13.45 ($p=0.004$) & 4.00$^\ast$& 1.00$^\ast$& 1.00 & 4.00 \\
C8 | Habit/interest & 1.92 ($p=0.59$) & 4.00 & 3.08 & 4.00&4.00\\
C9 | Illegal/inappropriate content & 12.75 ($p=0.005$) & 4.00$^\ast$ & 1.00$^\ast$$^\alpha$& 4.00 & 4.00$^\alpha$ \\C10 |Personal moment & 14.18 ($p=0.005$) & 4.00& 2.75 & 2.00 & 4.00 \\
\bottomrule

\end{tabular}
}
\caption{Median preference ratings for filtering techniques by image categories - matching symbols such as $\ast$ or $\alpha$ indicate statistically significant differences}
\Description{The table titled "Preference Ratings for Filtering Techniques by Image Category" shows how different visual filtering methods were rated in terms of comfort across ten image categories (C1–C10). The methods compared are Blurring, Pixelation, Content-Based Filling, and Masking. The table includes Friedman chi-squared test statistics and p-values to indicate statistical significance for each category. Asterisks (*) and matching Greek letters ( α) denote statistically significant differences between filtering techniques. For example, in Category 3 (C3), Blurring and Pixelation differ significantly; in C6 and C7, Blurring is significantly preferred over Pixelation.}
\label{filtering-preferences}
\end{table*}

\subsection{Target Audience}
Sensitivity while sharing images to family, friends and strangers can be highly subjective, with different individuals having varying levels of tolerance for specific content. When sharing potentially sensitive content, exercising caution and respect is essential. The comfort of sharing image levels may differ depending on the audience and context. In our study, we identified three target audiences for the BVIs: Family, Friends, and Strangers - a similar audiences were also used in \cite{akter2020uncomfortable}. 
These categories were selected to capture a range of interpersonal closeness and social boundaries commonly found in everyday sharing contexts. These audience categories were also used in technology acceptance models \cite{ahlstrom2014you,rico2010gesture, koelle2020social}. 
\textit{Family} represents a trusted circle with strong emotional bonds, where participants may feel more secure sharing personal or sensitive images. \textit{Friends} encompass a broader yet still personally connected group, often characterized by shared experiences and mutual trust. \textit{Strangers}, on the other hand, represent the least intimate and most socially distant category, where individuals are more likely to experience discomfort or hesitancy in sharing private content. This tripartite classification provides a practical and theoretically grounded framework for examining comfort levels in privacy-sensitive image sharing across varying degrees of familiarity and social exposure.
We anticipate that BVI participants will likely feel more comfortable sharing certain types of images with Family or Friends rather than Strangers.

% \subsection{Image data-set and description} 
\subsection{Participants}

% https://dl.acm.org/doi/pdf/10.1145/2702123.2702334
We recruited 20 BVI participants, ages 20 to 31 (mean age = 24.15, SD = 2.96), of whom 7 were female. Table \ref{par} shows participants' demographics.  This research work is approved by the Institutional Ethics Review Board under ethics number H23-02020.
% please see Supplementary Material for participant details 
We encouraged diversity by setting inclusion criteria that welcomed participants of different genders, ages, degrees of visual impairment, and levels of experience with assistive technologies. The study was conducted online to remove mobility barriers. Recruitment was carried out through flyers, word-of-mouth, local BVI networks, including a non-profit organization and community mailing list, and participant selection was monitored to ensure representation across these characteristics. Participants received \$15 for completing the study.

\subsection{Procedure}
Individual online video conferencing sessions via Zoom were arranged for the study participants to ensure accessibility and participation from individuals across various parts of a large city, which would have been logistically challenging with in-person sessions. Prior to each session, participants received the necessary details, including the session date, time, meeting link, and study details, as well as a consent form. We conducted the sessions via the institutional Zoom where audio and video were recorded for transcription purposes. Each session took one hour to complete. Upon joining the video call, demographic information such as age, gender, location, and type of visual impairment was gathered from each participant. Researchers then outlined the study objectives, explained the tasks and procedures that participants would undertake. 
% For each participant, the order is as follows:?
% 1. show one image
% 2. rate how comfortable they are with sharing the image as is with others (family, friends, strangers)
% 3. Compare techniques (what is the order? counterbalanced?)
%    3.a describe the technique 1 <- what is the description, same as one derived in 2.1?
%    3.b ask for preference
%    3.c compare comfort sharing with others (family, friends, strangers) (what is the order? counterbalanced?)
%       3.c.1 for each group, ask about comfort

A set of 20 images was randomly presented to the participants. Each image was verbally described in detail to the BVI participants, which was derived and validated with the BVI HCI researcher involved in the project. Participants were asked to explain why they felt an image was embarrassing and to identify the specific parts of the image that contributed to this feeling. This was asked to ensure that participants can also identify the right part that could trigger embarrassment while sharing images. Participants then rated their comfort level of sharing each image with family, friends, and strangers on a scale of 1 to 7, where 1 indicated strong discomfort and 7 indicated strong comfort. Following this, participants evaluated their preference for each image filtering using a 7-point Likert scale for four techniques: masking, blurred, pixelated, and content-based filling, where 1 indicated the least preferred technique, while a rating of 7 indicated the most preferred. After that, participants were asked to rate their comfort level in sharing the filtered images (i.e., image after applying their preferred filtering method) with friends, family, and strangers on a 7-point Likert scale.
\section{Results}
\subsection{Preference Scores for Filtering Techniques}

We used a Friedman test to analyze preferences for photo-sharing filters across 10 categories. Table \ref{filtering-preferences} shows test results with the median preference rating for filtering techniques. Post-hoc pairwise comparisons were Bonferroni adjusted ($\alpha$-level 0.05 to 0.008).

We first discuss the image categories for which we observed a significant difference.
\textit{C3 (Nudity or partial nudity)} showed a significant effect ($\chi^2(3,20)=12.66, p=.005$), with Blurring preferred over Pixelation ($p=.001$). \textit{C4 (Private area at home)} was also significant ($\chi^2(3,20)=9.92, p=.019$), with Blurring preferred over Pixelation ($p=.003$). For \textit{C6 (Identity information)}, the Friedman test was significant ($\chi^2(3,20)=14.60, p=.002$), again showing a preference for Blurring over Pixelation ($p=.001$). \textit{C7 (Medical condition)} showed a significant effect ($\chi^2(3,20)=13.45, p=.004$), with Blurring preferred over Pixelation ($p=.004$). Finally, \textit{C9 (Illegal/inappropriate content)} revealed significant differences ($\chi^2(3,20)=12.75, p=.005$), where both Blurring ($p=.004$) and Masking ($p=.004$) were preferred over Pixelation.

No significant differences were found for \textit{C1 (The identity of people)}, \textit{C2 (Children)}, \textit{C5 (Embarrassing shots)}, or \textit{C8 (Habit/interest)}. Although \textit{C10 (Personal moment)} showed an overall effect, no pairwise comparisons reached significance. Please refer to the Table \ref{filtering-preferences} for $\chi^2, p$ and median preference ratings.
 
\begin{figure*}[!h]
    \centering    
    \includegraphics[width=1.0\textwidth]{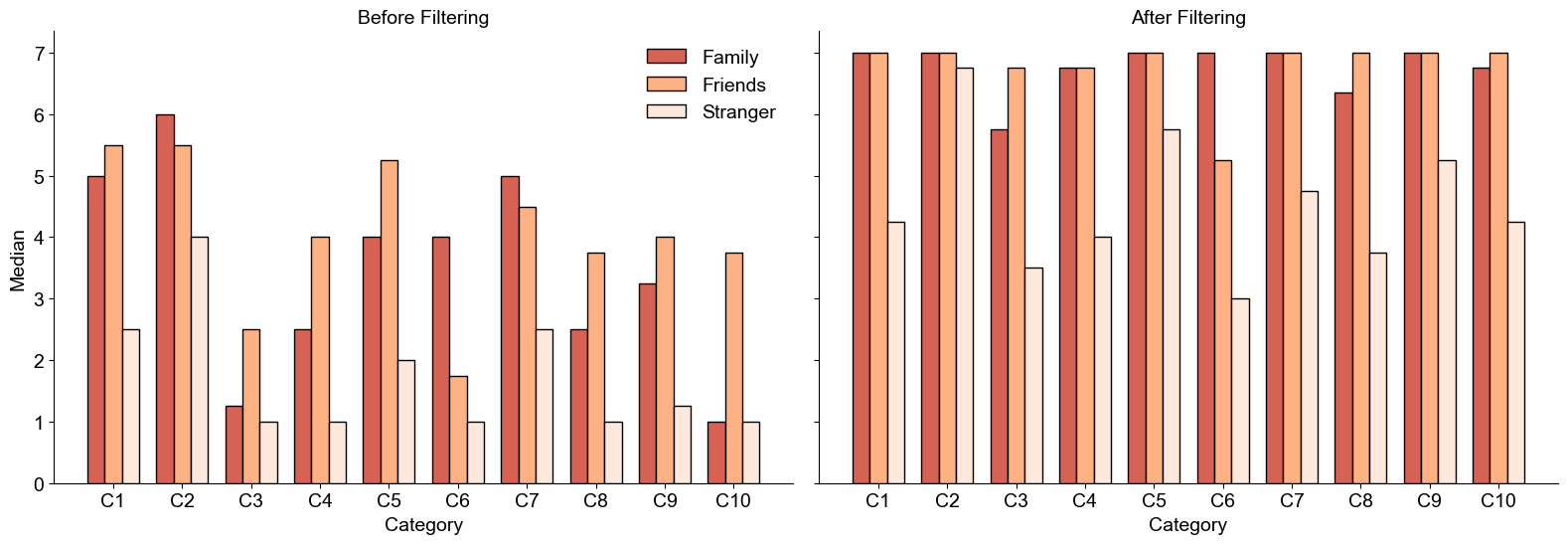}
    \caption{Bar plots comparing median values across ten categories for three audiences (Family, Friends, Stranger), before and after data filtering.}
    \Description{This figure contains two side-by-side bar plots. Each plot shows median values for ten categories (C1 to C10) grouped by audiences: Family, Friends, and Stranger. The left plot represents data before filtering, and the right plot represents data after filtering. Each group is shown in a distinct color. Across most categories, median values increase from the "Before Filtering" chart to the "After Filtering" chart. Family and Friends generally have higher medians than Strangers in both charts.}
    \label{before after bar chart}
\end{figure*}

\subsection{Preference Scores}

In addition, we analyzed how comfort levels varied by target audience - family, friends, and strangers - as shown in Figure \ref{before after bar chart}. To assess these differences, we again used a Friedman test followed by Wilcoxon tests across the ten image categories. Refer to Table~\ref{tab:combined-before-after} for test statistics ($\chi^2$, $p$) and median comfort ratings.

\textbf{Before applying filters:} For all ten image categories, test results showed significant differences in comfort ratings across \textit{Audiences} (Family, Friends, Strangers). 
Pairwise comparisons revealed a consistent pattern: all categories except for \textit{C10 (Personal Moment)}, participants were significantly more comfortable sharing images with \textit{Family} and \textit{Friends} than with \textit{Strangers}. For \textit{C10 (Personal Moment)}, participants' comfort ratings differed significantly between Family and Friends, and between Friends and Strangers.
All audience pairs (i.e., Family $\leftrightarrow$ Friends, Family $\leftrightarrow$ Strangers, Friends $\leftrightarrow$ Strangers) differences were significant for \textit{C5 (Embarrassing Shots)}, \textit{C6 (Identity Information)}, and \textit{C9 (Illegal/Inappropriate Content)}.

\textbf{After Applying Filters:} Across all ten categories, comfort ratings differed significantly across \textit{Audiences} (Family, Friends, Strangers). Pairwise comparisons showed a consistent pattern: 
For all categories, sharing images with both Family and Friends was rated significantly more comfortable than with Strangers. Additionally, all audience pairs (i.e., Family $\leftrightarrow$ Friends, Family $\leftrightarrow$ Strangers, Friends $\leftrightarrow$ Strangers) were significantly different for \textit{C3 (Nudity or Partial Nudity)} and \textit{C6 (Identity Information)}.

\begin{table*}[t]

\centering
\small
\resizebox{\textwidth}{!}{%
\begin{tabular}{l cccc c cccc}
\toprule
& \multicolumn{4}{c}{\textbf{Before Filtering}} & &
  \multicolumn{4}{c}{\textbf{After Filtering}} \\
\cmidrule(lr){2-5} \cmidrule(lr){7-10}
\textbf{Code | Category} &
$\chi^2$ (p) & Family & Friends & Stranger & &
$\chi^2$ (p) & Family & Friends & Stranger \\
\midrule

C1 | The identity of People &
25.90 ($<.001$) & 5.00$^\ast$ & 5.50$^\alpha$ & 2.50$^{\ast\alpha}$ &
& 20.84 ($<.001$) & 7.00$^\ast$ & 7.00$^\alpha$ & 4.25$^{\ast\alpha}$ \\

C2 | Children &
20.85 ($<.001$) & 6.00$^\ast$ & 5.50$^\alpha$ & 4.00$^{\ast\alpha}$ &
& 14.00 ($<.001$) & 7.00$^\ast$ & 7.00$^\alpha$ & 6.75$^{\ast\alpha}$ \\

C3 | Nudity or partial nudity &
20.17 ($<.001$) & 1.25$^\ast$ & 2.50$^\alpha$ & 1.00$^{\ast\alpha}$ &
& 20.17 ($<.001$) & 5.75$^{\ast\$}$ & 6.75 $^{\alpha\$}$ & 3.50$^{\ast\alpha}$ \\

C4 | Private area at home &
22.29 ($<.001$) & 2.50 $^\ast$ & 4.00$^\alpha$ & 1.00$^{\ast\alpha}$ &
& 21.28 ($<.001$) & 6.75 $^\ast$ & 6.75$^\alpha$ & 4.00$^{\ast\alpha}$ \\

C5 | Embarrassing shots &
16.54 ($<.001$) & 4.00$^{\ast\$}$ & 5.25$^{\alpha\$}$ & 2.00$^{\ast\alpha}$ &
& 16.15 ($<.001$) & 7.00$^\ast$ & 7.00$^\alpha$ & 5.75$^{\ast\alpha}$ \\

C6 | Identity information &
23.53 ($<.001$) & 4.00$^{\ast\$}$ & 1.75$^{\alpha\$}$ & 1.00$^{\ast\alpha}$ &
& 25.15 ($<.001$) & 7.00$^{\ast\$}$ & 5.25$^{\alpha\$}$ & 3.00$^{\ast\alpha}$ \\

C7 | Medical condition &
19.57 ($<.001$) & 5.00$^\ast$ & 4.50$^\alpha$ & 2.50$^{\ast\alpha}$ &
& 24.37 ($<.001$) & 7.00$^\ast$ & 7.00$^\alpha$ & 4.75$^{\ast\alpha}$ \\

C8 | Habit/interest &
21.93 ($<.001$) & 2.50$^\ast$ & 3.75$^\alpha$ & 1.00$^{\ast\alpha}$ &
& 20.28 ($<.001$) & 6.35$^\ast$ & 7.00$^\alpha$ & 3.75$^{\ast\alpha}$ \\

C9 | Illegal/inappropriate content &
25.80 ($<.001$) & 3.25$^{\ast\$}$ & 4.00$^{\ast\$}$ & 1.25$^{\alpha\$}$ &
& 21.26 ($<.001$) & 7.00$^\ast$ & 7.00$^\alpha$ & 5.25$^{\ast\alpha}$ \\

C10 | Personal moment &
21.78 ($<.001$) & 1.00$^\$$ & 3.75$^{\$\alpha}$ & 1.00$^\alpha$ &
& 19.51 ($<.001$) & 6.75$^\ast$ & 7.00$^\alpha$ & 4.25$^{\ast\alpha}$ \\

\bottomrule
\end{tabular}
}

\caption{Friedman test results and median preference scores for sharing with target Audiences — Before and After Applying Filters.  Here, \${}\ indicates a significant difference between Family and Friends,  $\alpha$ indicates a significant difference between Friends and Strangers,  and $\ast$ indicates a significant difference between Family and Strangers.}
\Description {
The table compares median comfort ratings for sharing images across Family, Friends, and Strangers for ten content categories, shown before and after applying filtering techniques. All categories show significant differences between audiences. Before filtering, comfort is generally moderate for Family and Friends and lowest for Strangers, especially for sensitive categories such as nudity, identity information, and medical content. After filtering, comfort increases across all audiences, with the largest improvements for Family and Friends, though Strangers consistently remain the lowest-rated group. The table also reports Friedman chi-square values, indicating that audience differences are statistically significant for every category. Significance markers denote which audience pairs differ.

}
\label{tab:combined-before-after}
\end{table*}

\subsection{Preference Scores Before/After Applying Filter}
Participants’ comfort levels in sharing images were assessed both before and after applying their most preferred obfuscation technique for each image category. We used Wilcoxon pairwise comparison to examine the significance of changes in comfort levels before and after applying the filter. The results, presented in Table \ref{friedman-by-audience}, show a statistically significant increase in comfort across all categories (i.e., from C1 to C10) and audience types (Family, Friends, and Stranger) after filtering was applied.

\begin{table*}[t]

\centering

\resizebox{\textwidth}{!}{%
\begin{tabular}{l c c c c c c}
\toprule
\multicolumn{1}{l}{\textbf{Code | Category}}
& \multicolumn{2}{c}{\textbf{Family}}
& \multicolumn{2}{c}{\textbf{Friends}}
& \multicolumn{2}{c}{\textbf{Stranger}} \\

\cmidrule(lr){2-3} \cmidrule(lr){4-5} \cmidrule(lr){6-7}
& $\chi^2$ (p) & Before $\to$ After
& $\chi^2$ (p) & Before $\to$ After
& $\chi^2$ (p) & Before $\to$ After \\
\midrule
C1 | The identity of people & 13.00 ($p<0.001$) & 5.00  $\to$  7.00 & 11.00 ($p<0.001$) & 5.50  $\to$  7.00 & 14.00 ($p<0.001$) & 2.50  $\to$  4.25 \\
C2 |  Children  & 10.00 ($p=0.002$) & 6.00  $\to$  7.00 & 8.33 ($p=0.004$) & 5.50  $\to$  7.00 & 14.00 ($p<0.001$) & 4.00  $\to$ 6.75 \\
C3 |  Nudity or partial nudity  & 13.24 ($p<0.001$) & 1.25  $\to$ 5.75 & 13.00 ($p<0.001$) & 2.50  $\to$ 6.75 & 14.00 ($p<0.001$) & 1.00  $\to$  3.50 \\
C4 |  Private area at home & 18.00 ($p<0.001$) & 2.50 $\to$  6.75 & 14.22 ($p<0.001$) & 4.00  $\to$ 6.75 & 17.00 ($p<0.001$) & 1.00  $\to$  4.00 \\
 C5 | Embarrassing shots & 14.22 ($p<0.001$) & 4.00  $\to$  7.00 & 13.24 ($p<0.001$) & 5.25  $\to$ 7.00 & 14.00 ($p<0.001$) & 2.00  $\to$  5.75 \\
C6 | Identity information  & 7.14 ($p=0.008$) & 4.00  $\to$  7.00 & 10.29 ($p=0.001$) & 1.75  $\to$  5.25 & 8.33 ($p=0.004$) & 1.00  $\to$  3.00 \\
 C7 |  Medical condition & 12.00 ($p<0.001$) & 5.00  $\to$ 7.00 & 16.00 ($p<0.001$) & 4.50  $\to$  7.00 & 7.14 ($p=0.008$) & 2.50  $\to$  4.75 \\
C8
| Habit/interest  & 16.00 ($p<0.001$) & 2.50  $\to$  6.35 & 8.07 ($p=0.005$) & 3.75  $\to$ 7.00 & 14.00 ($p<0.001$) & 1.00  $\to$  3.75 \\
C9
|  Illegal/inappropriate content & 17.00 ($p<0.001$) & 3.25  $\to$  7.00 & 14.00 ($p<0.001$) & 4.00 $\to$  7.00 & 14.00 ($p<0.001$) & 1.25  $\to$  5.25 \\
 
C10 |  Personal moment & 18.00 ($p<0.001$) & 1.00  $\to$  6.75 & 17.00 ($p<0.001$) & 3.75  $\to$ 7.00 & 12.25 ($p<0.001$) & 1.00  $\to$  4.25 \\

\bottomrule
\end{tabular}
}
\caption{Friedman test results and median ratings before and after filtering by audience type. All the pairwise comparisons revealed significant differences.}
\Description{The table titled “Friedman Test Results and Median Ratings Before and After Filtering by Audience Type” displays statistical results across 10 image categories (C1–C10) for three types of audiences: Family, Friends, and Strangers. For each category and audience group, the table presents the Friedman test statistic (chi-squared and p-value), followed by median comfort ratings before and after applying visual filtering. All p-values indicate statistically significant differences, with improvements in comfort levels after filtering across all groups and categories. The ratings tend to be higher for familiar audiences like Family and Friends, but improvements are also notable for Strangers. This table supports the subsection: Preference Scores for Sharing with Target Audience.}
\label{friedman-by-audience}
\end{table*}

\subsection{Subjective Feedback}
Pixelation was consistently rated lower across all image categories, a pattern also reflected in participant feedback: ``Pixelation is less preferred'' [P3]. The blocky, artificial appearance of pixelation may conflict with BVI  users' expectations, as it can obscure important visual cues and reduce the overall clarity of images. Participants echoed this concern - ``I know if I change it, pixelation might kill the aesthetic'' [P11] and emphasized its limited usefulness when more targeted methods are available - ``As there is a part to remove, I will use content-based filling - pixelation is not needed as it can be blocky'' [P18]. Content-Based Filling received mixed ratings, performing well in some categories (e.g., C1, C4, C5) but poorly in others (e.g., C6, C7), suggesting that its effectiveness is highly context-dependent. Participants also highlighted that filtering preferences vary by image type, with no single technique universally preferred. As one participant noted ``For different pictures, different techniques are better'' [P2] and another stated that the filter choice ``depends on the content'' [P8].

While most participants preferred filtered over unfiltered images for sharing, particularly with strangers, even modified images were sometimes deemed too sensitive or personal (e.g., images under C3 category). This was often due to the nature of the content, individual beliefs, or concerns about misinterpretation. For instance, one participant
remarked, ``Even after blurring, I feel awkward to share this with anyone'' [P19] for images in C3 category. Another noted, ``Everyone has their own belief, so I won’t share with strangers - they may judge'' [P14]. Another commented,
``Modified pictures may seem odd to strangers.'' [P20] for C3 images. Together, these findings suggest that systems should offer multiple filtering options, personalization or adaptive filtering, possibly paired with intelligent recommendations based on content and recipient type (e.g., stranger vs. friend).

\section{DISCUSSION AND DESIGN GUIDELINES}

Our study offers insights into how BVI individuals evaluate image filtering techniques across sensitive image types and how filtering affects their comfort sharing images with different audiences. Unlike prior work focused on single filtering methods or specific content types \cite{hasan2018viewer}, our findings reveal a more refined relationship between image content, social context, and obfuscation preference. We next discuss these findings within privacy theory, analyze the implications for BVI-centered design, and outline directions for future work.

\textbf{Content Sensitivity Shapes Filtering Preferences} Consistent with prior research showing that obfuscation is not one-size-fits-all \cite{hasan2018viewer}, participants’ preferences varied across ten content categories. Pixelation - commonly used in mainstream media - was rated lowest, suggesting that its blocky distortion conflicts with BVI users’ expectations of how an image ``ought'' to be shared \cite{blurvsblock}. Blurring was often preferred for highly sensitive categories such as nudity or identity information, whereas masking and content-based filling were viewed as more appropriate when removing explicit details while preserving context. These findings extend prior work \cite{hasan2018viewer, fi17020055} by demonstrating that BVI users' judgments are content-dependent rather than technique-specific: they did not want a universal filter; instead, they wanted control to select filters that best fit the sensitive contents (answering RQ1). This aligns with privacy-by-design principles that emphasize user agency in configuring privacy protections \cite{cavoukian2009privacy}.

\textbf{Audience Context Strongly Influences Comfort} Participants constantly reported higher comfort sharing images with family and friends than with strangers. Importantly, filtering increased comfort levels across all audiences, but the largest gains were observed in the stranger condition (answering RQ2). This aligns with contextual integrity \cite{nissenbaum2004privacy}, which argues that privacy depends not only on what is shared but to whom information flows. Our results show that filtering helps restore contextual integrity by aligning image disclosure with social norms-especially when sensitive content risks being misunderstood or judged by strangers. This has important implications for the design of future VDS, where strangers or volunteers commonly interpret users’ images.

\textbf{Design Implications} Our findings translate into several design recommendations:

\begin{itemize}

\item \textit{Prioritize Context-Preserving Filters:} Pixelation was consistently rated poorly by BVI users, aligning with prior work showing that coarse distortions, e.g., heavy pixelation, reduce image utility without significantly improving privacy \cite{blurvsblock}. Instead of relying on pixelation, designers should prioritize filters that preserve spatial coherence and provide more usable representations, such as content-aware filling or selective blurring. This is particularly important as BVI users often rely on non-visual cues from VDS (audio descriptions, AI-generated summaries), which can be degraded by aggressive pixelation. Pixelation should therefore be treated as a last resort, not a default.

\item \textit{Design Flexible, User-Driven Personalization Mechanisms:} Offering multiple filtering options is helpful, but without thoughtful design, it risks shifting labor onto BVI users - an issue well documented in accessibility research \cite{das2019doesn}.
Systems should therefore support configurable, persistent personalization - such as allowing users to define rules like ``auto-blur identity information in images'' or ``auto-mask illegal content'' - to reduce repeated decision-making. 
These rules must remain transparent and reversible —allowing users to have full control over them. 
Drawing on consent and personalization frameworks \cite{knijnenburg2022user}, this approach provides meaningful customization without requiring constant effort from the user.

\item \textit{Implement Intelligent, Content and Audience-Aware Filtering:} Privacy for BVI users is highly dependent on social context - a finding supported by our study and prior work on contextual integrity \cite{nissenbaum2004privacy} and privacy practices \cite{Tousif}. Thus, building on existing selective-obfuscation tools \cite{zhang2023imageally}, we suggest that obfuscation systems for BVI users should automatically suggest filters for the intended audience, e.g., dynamically adjust obfuscation strength, flagging disclosure risks, or offer one-click sharing modes for varying audiences that apply appropriate filters proactively. 
This approach advances the field towards an adaptive, anticipatory model where obfuscation is socially intelligent.
\end{itemize}

\section{Limitations and Future Work}

Our image dataset includes only ten categories, each with two sample images, which restricts the diversity of visual materials and may not fully capture the range of sensitive content encountered in real-world scenarios. Although this allowed us to keep the study manageable for participants, a broader set of categories and a more diverse image pool would offer a richer understanding of how obfuscation preferences vary across different types of sensitive content. Future work should explore a broader range of scenarios and contexts, particularly those culturally or personally sensitive. We focused on comfort and willingness to share; our study did not directly assess image receiver (i.e., friends, family, or strangers who receive images from BVI after applying filters) participants’ ability to interpret filtered images, specifically, whether they could still extract the necessary information or assess the image’s aesthetics after obfuscation. Since BVI users often rely on assistive tools like alt text or descriptions, future research should investigate how filtering affects their comprehension and the practical usefulness of images. Understanding the trade-off between obfuscation and information loss is crucial for designing more effective and user-centered privacy tools. 
Our findings underscore the nuanced relationship between filtering techniques and social dynamics --- while filters can enhance comfort, they cannot fully override deeply personal, cultural, or relational considerations (e.g., content in category C3 - Nudity or partial nudity). Future research could explore considering these complexities in mind, allowing users to express their sharing intentions and adapt the filtering process accordingly. This could involve integrating personalization features, recipient-aware filtering, or user-driven adjustments that better reflect individual preferences and social context. During the study, we assessed participants’ comprehension of the techniques and encouraged them to ask clarifying questions. We proceeded only after they confirmed their understanding. However, we did not explicitly ask whether participants currently use similar techniques in their everyday image-sharing practices. We acknowledge that this may have influenced the results. Our study focused on comfort and willingness to share images, but did not examine how filtering affects their utility for supporting daily tasks through Visual Description Services. Participants' preferences may differ when images are evaluated for task effectiveness - such as how well each technique hides sensitive content - rather than comfort with sharing alone. As such, our findings are limited to comfort-based judgments, and future work should assess utility in real task contexts. Finally, participants were recruited from a metropolitan city, which limits international and cultural diversity. Expanding recruitment to include participants from non-metropolitan and international contexts would enhance the generalizability of future findings.

\section{Conclusion}
In this paper, we explore BVI individuals' privacy concerns when sharing images, particularly with varying target audiences - family, friends, and strangers. We identified preferences that vary by context and content type by exploring four different image filtering techniques, emphasizing the need for flexible, user-driven tools. While no single method proved universally preferred, participants felt more comfortable sharing filtered images to different audiences. 
Our findings underscore the importance of designing customizable and privacy-aware image-sharing solutions for BVI users.

\begin{acks}
This work was funded by NSERC Discovery Grant (\#RGPIN-2019- 05211), SSHRC Insight Development Grants (\#430-2025-01369) and the Canada Foundation for Innovation Infrastructure Fund (\#40440). Arshad Nasser was involved in the project during his affiliation with the University of British Columbia and is currently affiliated with King Fahd University of Petroleum \& Minerals.
\end{acks}
\bibliographystyle{ACM-Reference-Format.bst}
\bibliography{main.bib}

@inproceedings{10.1145/3281505.3281541,
author = {Alallah, Fouad and Neshati, Ali and Sakamoto, Yumiko and Hasan, Khalad and Lank, Edward and Bunt, Andrea and Irani, Pourang},
title = {Performer vs. observer: whose comfort level should we consider when examining the social acceptability of input modalities for head-worn display?},
year = {2018},
isbn = {9781450360869},
publisher = {Association for Computing Machinery},
address = {New York, NY, USA},
url = {https://doi.org/10.1145/3281505.3281541},
doi = {10.1145/3281505.3281541},
booktitle = {Proceedings of the 24th ACM Symposium on Virtual Reality Software and Technology},
articleno = {10},
numpages = {9},
location = {Tokyo, Japan},
series = {VRST '18}
}

@inproceedings{10.1145/3411764.3445430,
author = {Pandey, Laxmi and Hasan, Khalad and Arif, Ahmed Sabbir},
title = {Acceptability of Speech and Silent Speech Input Methods in Private and Public},
year = {2021},
isbn = {9781450380966},
publisher = {Association for Computing Machinery},
address = {New York, NY, USA},
url = {https://doi.org/10.1145/3411764.3445430},
doi = {10.1145/3411764.3445430},
booktitle = {Proceedings of the 2021 CHI Conference on Human Factors in Computing Systems},
articleno = {251},
numpages = {13},
location = {Yokohama, Japan},
series = {CHI '21}
}

@article{zhao2017effect,
  title={The effect of computer-generated descriptions on photo-sharing experiences of people with visual impairments},
  author={Zhao, Yuhang and Wu, Shaomei and Reynolds, Lindsay and Azenkot, Shiri},
  journal={Proceedings of the ACM on Human-Computer Interaction},
  volume={1},
  number={CSCW},
  pages={1--22},
  year={2017},
  publisher={ACM New York, NY, USA}
}

@inproceedings{rico2010gesture,
  title={Gesture and voice prototyping for early evaluations of social acceptability in multimodal interfaces},
  author={Rico, Julie and Brewster, Stephen},
  booktitle={International Conference on Multimodal Interfaces and the Workshop on Machine Learning for Multimodal Interaction},
  pages={1--9},
  year={2010}
}

@inproceedings{10.1145/2070481.2070551,
author = {Williamson, Julie R. and Crossan, Andrew and Brewster, Stephen},
title = {Multimodal mobile interactions: usability studies in real world settings},
year = {2011},
isbn = {9781450306416},
publisher = {Association for Computing Machinery},
address = {New York, NY, USA},
url = {https://doi.org/10.1145/2070481.2070551},
doi = {10.1145/2070481.2070551},
booktitle = {Proceedings of the 13th International Conference on Multimodal Interfaces},
pages = {361–368},
numpages = {8},
location = {Alicante, Spain},
series = {ICMI '11}
}

@inproceedings{10.1145/1851600.1851647,
author = {Montero, Calkin S. and Alexander, Jason and Marshall, Mark T. and Subramanian, Sriram},
title = {Would you do that? understanding social acceptance of gestural interfaces},
year = {2010},
isbn = {9781605588353},
publisher = {Association for Computing Machinery},
address = {New York, NY, USA},
url = {https://doi.org/10.1145/1851600.1851647},
doi = {10.1145/1851600.1851647},
booktitle = {Proceedings of the 12th International Conference on Human Computer Interaction with Mobile Devices and Services},
pages = {275–278},
numpages = {4},
location = {Lisbon, Portugal},
series = {MobileHCI '10}
}

@article{nissenbaum2004privacy,
  title={Privacy as contextual integrity},
  author={Nissenbaum, Helen},
  journal={Wash. L. Rev.},
  volume={79},
  pages={119},
  year={2004},
  publisher={HeinOnline}
}

@article{cavoukian2009privacy,
  title={Privacy by design: The 7 foundational principles},
  author={Cavoukian, Ann and others},
  journal={Information and privacy commissioner of Ontario, Canada},
  volume={5},
  number={2009},
  pages={12},
  year={2009}
}

@article{das2019doesn,
  title={" It doesn't win you friends" Understanding Accessibility in Collaborative Writing for People with Vision Impairments},
  author={Das, Maitraye and Gergle, Darren and Piper, Anne Marie},
  journal={Proceedings of the ACM on Human-Computer Interaction},
  volume={3},
  number={CSCW},
  pages={1--26},
  year={2019},
  publisher={ACM New York, NY, USA}
}

@Article{fi17020055,
AUTHOR = {Cruz, Ana Cassia and Costa, Rogério Luís de C. and Santos, Leonel and Rabadão, Carlos and Marto, Anabela and Gonçalves, Alexandrino},
TITLE = {Assessing User Perceptions and Preferences on Applying Obfuscation Techniques for Privacy Protection in Augmented Reality},
JOURNAL = {Future Internet},
VOLUME = {17},
YEAR = {2025},
NUMBER = {2},
ARTICLE-NUMBER = {55},
URL = {https://www.mdpi.com/1999-5903/17/2/55},
ISSN = {1999-5903},
DOI = {10.3390/fi17020055}
}

@inproceedings{ahlstrom2014you,
  title={Are you comfortable doing that? Acceptance studies of around-device gestures in and for public settings},
  author={Ahlstr{\"o}m, David and Hasan, Khalad and Irani, Pourang},
  booktitle={Proceedings of the 16th international conference on Human-computer interaction with mobile devices \& services},
  pages={193--202},
  year={2014}
}

@inproceedings{koelle2020social,
  title={Social acceptability in HCI: A survey of methods, measures, and design strategies},
  author={Koelle, Marion and Ananthanarayan, Swamy and Boll, Susanne},
  booktitle={Proceedings of the 2020 CHI Conference on Human Factors in Computing Systems},
  pages={1--19},
  year={2020}
}

@inproceedings{hasan2018viewer,
  title={Viewer experience of obscuring scene elements in photos to enhance privacy},
  author={Hasan, Rakibul and Hassan, Eman and Li, Yifang and Caine, Kelly and Crandall, David J and Hoyle, Roberto and Kapadia, Apu},
  booktitle={Proceedings of the 2018 CHI Conference on Human Factors in Computing Systems},
  pages={1--13},
  year={2018}
}

@article{padilla2015visual,
  title={Visual privacy by context: proposal and evaluation of a level-based visualisation scheme},
  author={Padilla-L{\'o}pez, Jos{\'e} Ram{\'o}n and Chaaraoui, Alexandros Andre and Gu, Feng and Fl{\'o}rez-Revuelta, Francisco},
  journal={Sensors},
  volume={15},
  number={6},
  pages={12959--12982},
  year={2015},
  publisher={MDPI}
}

@inproceedings{bragg2020exploring,
  title={Exploring collection of sign language datasets: Privacy, participation, and model performance},
  author={Bragg, Danielle and Koller, Oscar and Caselli, Naomi and Thies, William},
  booktitle={Proceedings of the 22nd International ACM SIGACCESS Conference on Computers and Accessibility},
  pages={1--14},
  year={2020}
}

@inproceedings{shu2018cardea,
  title={Cardea: Context-aware visual privacy protection for photo taking and sharing},
  author={Shu, Jiayu and Zheng, Rui and Hui, Pan},
  booktitle={Proceedings of the 9th ACM Multimedia Systems Conference},
  pages={304--315},
  year={2018}
}

@inproceedings{ilia2015face,
  title={Face/off: Preventing privacy leakage from photos in social networks},
  author={Ilia, Panagiotis and Polakis, Iasonas and Athanasopoulos, Elias and Maggi, Federico and Ioannidis, Sotiris},
  booktitle={Proceedings of the 22nd ACM SIGSAC Conference on computer and communications security},
  pages={781--792},
  year={2015}
}

@book{brunton2015obfuscation,
  title={Obfuscation: A user's guide for privacy and protest},
  author={Brunton, Finn and Nissenbaum, Helen},
  year={2015},
  publisher={Mit Press}
}

@inproceedings{voykinska2016blind,
  title={How blind people interact with visual content on social networking services},
  author={Voykinska, Violeta and Azenkot, Shiri and Wu, Shaomei and Leshed, Gilly},
  booktitle={Proceedings of the 19th acm conference on computer-supported cooperative work \& social computing},
  pages={1584--1595},
  year={2016}
}

@inproceedings{stangl2023dump,
  title={“Dump it, Destroy it, Send it to Data Heaven”: Blind People’s Expectations for Visual Privacy in Visual Assistance Technologies},
  author={Stangl, Abigale and Sadjo, Emma and Emami-Naeini, Pardis and Wang, Yang and Gurari, Danna and Findlater, Leah},
  booktitle={Proceedings of the 20th International Web for All Conference},
  pages={134--147},
  year={2023}
}

@article{alharbi2022understanding,
  title={Understanding emerging obfuscation technologies in visual description services for blind and low vision people},
  author={Alharbi, Rahaf and Brewer, Robin N and Schoenebeck, Sarita},
  journal={Proceedings of the ACM on Human-Computer Interaction},
  volume={6},
  number={CSCW2},
  pages={1--33},
  year={2022},
  publisher={ACM New York, NY, USA}
}

@inproceedings{zhang2023imageally,
  title={$\{$ImageAlly$\}$: A $\{$Human-AI$\}$ Hybrid Approach to Support Blind People in Detecting and Redacting Private Image Content},
  author={Zhang, Zhuohao Jerry and Kaushik, Smirity and Seo, JooYoung and Yuan, Haolin and Das, Sauvik and Findlater, Leah and Gurari, Danna and Stangl, Abigale and Wang, Yang},
  booktitle={Nineteenth Symposium on Usable Privacy and Security (SOUPS 2023)},
  pages={417--436},
  year={2023}
}

@inproceedings{zhang2024designing,
  title={Designing Accessible Obfuscation Support for Blind Individuals’ Visual Privacy Management},
  author={Zhang, Lotus and Stangl, Abigale and Sharma, Tanusree and Tseng, Yu-Yun and Xu, Inan and Gurari, Danna and Wang, Yang and Findlater, Leah},
  booktitle={Proceedings of the 2024 CHI Conference on Human Factors in Computing Systems},
  pages={1--19},
  year={2024}
}

@inproceedings{bennett2018teens,
  title={How teens with visual impairments take, edit, and share photos on social media},
  author={Bennett, Cynthia L and E, Jane and Mott, Martez E and Cutrell, Edward and Morris, Meredith Ringel},
  booktitle={Proceedings of the 2018 CHI conference on human factors in computing systems},
  pages={1--12},
  year={2018}
}

@inproceedings{morris2016most,
  title={" With most of it being pictures now, I rarely use it" Understanding Twitter's Evolving Accessibility to Blind Users},
  author={Morris, Meredith Ringel and Zolyomi, Annuska and Yao, Catherine and Bahram, Sina and Bigham, Jeffrey P and Kane, Shaun K},
  booktitle={Proceedings of the 2016 CHI conference on human factors in computing systems},
  pages={5506--5516},
  year={2016}
}

@inproceedings{orekondy2017towards,
  title={Towards a visual privacy advisor: Understanding and predicting privacy risks in images},
  author={Orekondy, Tribhuvanesh and Schiele, Bernt and Fritz, Mario},
  booktitle={Proceedings of the IEEE international conference on computer vision},
  pages={3686--3695},
  year={2017}
}

@inproceedings{besmer2010moving,
  title={Moving beyond untagging: photo privacy in a tagged world},
  author={Besmer, Andrew and Richter Lipford, Heather},
  booktitle={Proceedings of the SIGCHI Conference on Human Factors in Computing Systems},
  pages={1563--1572},
  year={2010}
}

@inproceedings{aura2006scanning,
  title={Scanning electronic documents for personally identifiable information},
  author={Aura, Tuomas and Kuhn, Thomas A and Roe, Michael},
  booktitle={Proceedings of the 5th ACM workshop on Privacy in electronic society},
  pages={41--50},
  year={2006}
}

@inproceedings{hoyle2015sensitive,
  title={Sensitive lifelogs: A privacy analysis of photos from wearable cameras},
  author={Hoyle, Roberto and Templeman, Robert and Anthony, Denise and Crandall, David and Kapadia, Apu},
  booktitle={Proceedings of the 33rd Annual ACM conference on human factors in computing systems},
  pages={1645--1648},
  year={2015}
}

@article{adams2007sharing,
  title={Sharing, privacy and trust issues for photo collections},
  author={Adams, Anne and Cunningham, Sally Jo and Masoodian, Masood},
  year={2007},
  publisher={University of Waikato}
}

@inproceedings{10.1145/3613904.3642030,
author = {Xie, Jingyi and Yu, Rui and Zhang, He and Lee, Sooyeon and Billah, Syed Masum and Carroll, John M.},
title = {BubbleCam: Engaging Privacy in Remote Sighted Assistance},
year = {2024},
isbn = {9798400703300},
publisher = {Association for Computing Machinery},
address = {New York, NY, USA},
url = {https://doi.org/10.1145/3613904.3642030},
doi = {10.1145/3613904.3642030},
booktitle = {Proceedings of the 2024 CHI Conference on Human Factors in Computing Systems},
articleno = {48},
numpages = {16},
location = {Honolulu, HI, USA},
series = {CHI '24}
}

@inproceedings{10.1145/3663548.3675599,
author = {Chang, Ruei-Che and Liu, Yuxuan and Zhang, Lotus and Guo, Anhong},
title = {EditScribe: Non-Visual Image Editing with Natural Language Verification Loops},
year = {2024},
isbn = {9798400706776},
publisher = {Association for Computing Machinery},
address = {New York, NY, USA},
url = {https://doi.org/10.1145/3663548.3675599},
doi = {10.1145/3663548.3675599},
booktitle = {Proceedings of the 26th International ACM SIGACCESS Conference on Computers and Accessibility},
articleno = {65},
numpages = {19},
location = {St. John's, NL, Canada},
series = {ASSETS '24}
}

@inproceedings{10.1145/3663548.3675612,
author = {Kamikubo, Rie and Zamiri Zeraati, Farnaz and Lee, Kyungjun and Kacorri, Hernisa},
title = {AccessShare: Co-designing Data Access and Sharing with Blind People},
year = {2024},
isbn = {9798400706776},
publisher = {Association for Computing Machinery},
address = {New York, NY, USA},
url = {https://doi.org/10.1145/3663548.3675612},
doi = {10.1145/3663548.3675612},
booktitle = {Proceedings of the 26th International ACM SIGACCESS Conference on Computers and Accessibility},
articleno = {60},
numpages = {16},
location = {St. John's, NL, Canada},
series = {ASSETS '24}
}

@inproceedings{10.1145/3663548.3675635,
author = {Hong, Jonggi and Kacorri, Hernisa},
title = {Understanding How Blind Users Handle Object Recognition Errors: Strategies and Challenges},
year = {2024},
isbn = {9798400706776},
publisher = {Association for Computing Machinery},
address = {New York, NY, USA},
url = {https://doi.org/10.1145/3663548.3675635},
doi = {10.1145/3663548.3675635},
booktitle = {Proceedings of the 26th International ACM SIGACCESS Conference on Computers and Accessibility},
articleno = {63},
numpages = {15},
location = {St. John's, NL, Canada},
series = {ASSETS '24}
}

@article{c,
  title={Image inpainting: A review},
  author={Elharrouss, Omar and Almaadeed, Noor and Al-Maadeed, Somaya and Akbari, Younes},
  journal={Neural Processing Letters},
  volume={51},
  pages={2007--2028},
  year={2020},
  publisher={Springer}
}

@INPROCEEDINGS{f,
  author={Xue, Hanyu and Liu, Bo and Din, Ming and Song, Li and Zhu, Tianqing},
  booktitle={ICC 2020 - 2020 IEEE International Conference on Communications (ICC)}, 
  title={Hiding Private Information in Images From AI}, 
  year={2020},
  volume={},
  number={},
  pages={1-6},
  doi={10.1109/ICC40277.2020.9148656}}

@article{h,
  title={Web challenges faced by blind and vision impaired users in libraries of Delhi: An Indian scenario},
  author={Kumar, Shailendra and Sanaman, Gareema},
  journal={The Electronic Library},
  volume={33},
  number={2},
  pages={242--257},
  year={2015},
  publisher={Emerald Group Publishing Limited}
}

@inproceedings{i,
author = {Brady, Erin and Morris, Meredith Ringel and Zhong, Yu and White, Samuel and Bigham, Jeffrey P.},
title = {Visual challenges in the everyday lives of blind people},
year = {2013},
isbn = {9781450318990},
publisher = {Association for Computing Machinery},
address = {New York, NY, USA},
url = {https://doi.org/10.1145/2470654.2481291},
doi = {10.1145/2470654.2481291},
booktitle = {Proceedings of the SIGCHI Conference on Human Factors in Computing Systems},
pages = {2117–2126},
numpages = {10},
location = {Paris, France},
series = {CHI '13}
}

@inproceedings{j,
  title={Grounding answers for visual questions asked by visually impaired people},
  author={Chen, Chongyan and Anjum, Samreen and Gurari, Danna},
  booktitle={Proceedings of the IEEE/CVF Conference on Computer Vision and Pattern Recognition},
  pages={19098--19107},
  year={2022}
}

@ARTICLE{k,
  author={Newton, E.M. and Sweeney, L. and Malin, B.},
  journal={IEEE Transactions on Knowledge and Data Engineering}, 
  title={Preserving privacy by de-identifying face images}, 
  year={2005},
  volume={17},
  number={2},
  pages={232-243},
  doi={10.1109/TKDE.2005.32}}

@ARTICLE{l,
  author={Steinberger, Markus and Waldner, Manuela and Streit, Marc and Lex, Alexander and Schmalstieg, Dieter},
  journal={IEEE Transactions on Visualization and Computer Graphics}, 
  title={Context-Preserving Visual Links}, 
  year={2011},
  volume={17},
  number={12},
  pages={2249-2258},
  doi={10.1109/TVCG.2011.183}}

@inproceedings{data,
  title={Human perceptions of sensitive content in photos},
  author={Li, Yifang and Troutman, Wyatt and Knijnenburg, Bart P and Caine, Kelly},
  booktitle={Proceedings of the IEEE Conference on Computer Vision and Pattern Recognition Workshops},
  pages={1590--1596},
  year={2018}
}

@article{kuriakose2022tools,
  title={Tools and technologies for blind and visually impaired navigation support: a review},
  author={Kuriakose, Bineeth and Shrestha, Raju and Sandnes, Frode Eika},
  journal={IETE Technical Review},
  volume={39},
  number={1},
  pages={3--18},
  year={2022},
  publisher={Taylor \& Francis}
}

@inproceedings{akter2020uncomfortable,
  title={" I am uncomfortable sharing what I can't see": Privacy Concerns of the Visually Impaired with Camera Based Assistive Applications},
  author={Akter, Taslima and Dosono, Bryan and Ahmed, Tousif and Kapadia, Apu and Semaan, Bryan},
  booktitle={29th USENIX Security Symposium (USENIX Security 20)},
  pages={1929--1948},
  year={2020}
}

@misc{BeMyEyes,
  author       = {{Be My Eyes}},
  title        = {Be My Eyes - See the world together},
  year         = 2025,
  url          = {https://www.bemyeyes.com/},
  note         = {Accessed: 2025-02-02}
}

@misc{SeeingAI,
  author       = {{Seeing AI}},
  title        = {Seeing AI - Talking Camera App for the Blind},
  year         = 2025,
  url          = {https://www.seeingai.com/},
  note         = {Accessed: 2025-02-02}
}

@article{15,
author = {Alharbi, Rawan and Tolba, Mariam and Petito, Lucia C. and Hester, Josiah and Alshurafa, Nabil},
title = {To Mask or Not to Mask? Balancing Privacy with Visual Confirmation Utility in Activity-Oriented Wearable Cameras},
year = {2019},
issue_date = {September 2019},
publisher = {Association for Computing Machinery},
address = {New York, NY, USA},
volume = {3},
number = {3},
url = {https://doi.org/10.1145/3351230},
doi = {10.1145/3351230},
journal = {Proc. ACM Interact. Mob. Wearable Ubiquitous Technol.},
month = {sep},
articleno = {72},
numpages = {29}
}

@InProceedings{66,
author = {Gurari, Danna and Li, Qing and Lin, Chi and Zhao, Yinan and Guo, Anhong and Stangl, Abigale and Bigham, Jeffrey P.},
title = {VizWiz-Priv: A Dataset for Recognizing the Presence and Purpose of Private Visual Information in Images Taken by Blind People},
booktitle = {Proceedings of the IEEE/CVF Conference on Computer Vision and Pattern Recognition (CVPR)},
month = {June},
year = {2019}
}

@INPROCEEDINGS{blurvsblock,
  author={Li, Yifang and Vishwamitra, Nishant and Knijnenburg, Bart P. and Hu, Hongxin and Caine, Kelly},
  booktitle={2017 IEEE Conference on Computer Vision and Pattern Recognition Workshops (CVPRW)}, 
  title={Blur vs. Block: Investigating the Effectiveness of Privacy-Enhancing Obfuscation for Images}, 
  year={2017},
  volume={},
  number={},
  pages={1343-1351},
  doi={10.1109/CVPRW.2017.176}}

@article{90,
author = {Li, Yifang and Vishwamitra, Nishant and Knijnenburg, Bart P. and Hu, Hongxin and Caine, Kelly},
title = {Effectiveness and Users' Experience of Obfuscation as a Privacy-Enhancing Technology for Sharing Photos},
year = {2017},
issue_date = {November 2017},
publisher = {Association for Computing Machinery},
address = {New York, NY, USA},
volume = {1},
number = {CSCW},
url = {https://doi.org/10.1145/3134702},
doi = {10.1145/3134702},
journal = {Proc. ACM Hum.-Comput. Interact.},
month = {dec},
articleno = {67},
numpages = {24}
}

@inproceedings{10.1145/1866029.1866080,
author = {Bigham, Jeffrey P. and Jayant, Chandrika and Ji, Hanjie and Little, Greg and Miller, Andrew and Miller, Robert C. and Miller, Robin and Tatarowicz, Aubrey and White, Brandyn and White, Samual and Yeh, Tom},
title = {VizWiz: nearly real-time answers to visual questions},
year = {2010},
isbn = {9781450302715},
publisher = {Association for Computing Machinery},
address = {New York, NY, USA},
url = {https://doi.org/10.1145/1866029.1866080},
doi = {10.1145/1866029.1866080},
booktitle = {Proceedings of the 23nd Annual ACM Symposium on User Interface Software and Technology},
pages = {333–342},
numpages = {10},
location = {New York, New York, USA},
series = {UIST '10}
}

@misc{knijnenburg2022user,
  title={User-Tailored Privacy.},
  author={Knijnenburg, Bart P and Anaraky, Reza Ghaiumy and Wilkinson, Daricia and Namara, Moses and He, Yangyang and Cherry, David and Ash, Erin},
  year={2022}
}

@inproceedings{Tousif,
author = {Ahmed, Tousif and Hoyle, Roberto and Connelly, Kay and Crandall, David and Kapadia, Apu},
title = {Privacy Concerns and Behaviors of People with Visual Impairments},
year = {2015},
isbn = {9781450331456},
publisher = {Association for Computing Machinery},
address = {New York, NY, USA},
url = {https://doi.org/10.1145/2702123.2702334},
doi = {10.1145/2702123.2702334},
booktitle = {Proceedings of the 33rd Annual ACM Conference on Human Factors in Computing Systems},
pages = {3523–3532},
numpages = {10},
location = {Seoul, Republic of Korea},
series = {CHI '15}
}

@inproceedings{kacorri2017people,
  title={People with visual impairment training personal object recognizers: Feasibility and challenges},
  author={Kacorri, Hernisa and Kitani, Kris M and Bigham, Jeffrey P and Asakawa, Chieko},
  booktitle={Proceedings of the 2017 CHI Conference on Human Factors in Computing Systems},
  pages={5839--5849},
  year={2017}
}

@INPROCEEDINGS{4067752,
  author={Lanigan, Patrick E. and Paulos, Aaron M. and Williams, Andrew W. and Rossi, Dan and Narasimhan, Priya},
  booktitle={2006 10th IEEE International Symposium on Wearable Computers}, 
  title={Trinetra: Assistive Technologies for Grocery Shopping for the Blind}, 
  year={2006},
  volume={},
  number={},
  pages={147-148},
  doi={10.1109/ISWC.2006.286369}}

@misc{aira,
  title = {Aira App},
  author = {{Aira}},
  howpublished = {\url{https://aira.io/aira-app/}},
  year = {2025},
  note = {Accessed: 2025-02-02}
}

\end{document}